\documentclass[floatfix,a4paper,aps,prd,preprintnumbers,twocolumn]{revtex4-1} 
 \usepackage{amsmath} 
 \usepackage{amssymb} 
 \usepackage{amsfonts} 
 \usepackage{amscd} 
 \usepackage{accents} 
 \usepackage{bbm} 
 \usepackage{pst-all} 
 \usepackage{geometry} 
 \usepackage{epsfig} 
 \usepackage{slashed} 

 \newcommand{\bmat}{\left(\begin{array}} 
 \newcommand{\emat}{\end{array}\right)} 
 \newcommand{\be}{\begin{equation}} 
 \newcommand{\ee}{\end{equation}} 
 \newcommand{\ba}{\begin{eqnarray}} 
 \newcommand{\ea}{\end{eqnarray}}

 \def\lsim{\raise0.3ex\hbox{$\;<$\kern-0.75em\raise-1.1ex\hbox{$\sim\;$}}} 
 \def\gsim{\raise0.3ex\hbox{$\;>$\kern-0.75em\raise-1.1ex\hbox{$\sim\;$}}}

 \def\be{\beta}

 \newcommand{\bea}{\begin{eqnarray}}  
 \newcommand{\eea}{\end{eqnarray}}  
 \newcommand{\beq}{\begin{equation}}  
 \newcommand{\eeq}{\end{equation}}   
 \newcommand{\beqa}{\begin{eqnarray}}  
 \newcommand{\eeqa}{\end{eqnarray}}  
 \newcommand{\bit}{\begin{itemize}}  
 \newcommand{\eit}{\end{itemize}}  
 \newcommand{\barr}{\begin{eqnarray}} 
 \newcommand{\earr}{\end{eqnarray}}

 \newcommand{\met}{\ensuremath{\slashed{p}_T}}

 \newcommand{\spot}{\mathcal{W}} 
  
 \newcommand{\rp}{R_p} 
 \newcommand{\bt}{\textrm{B}_3} 
  
 \newcommand{\ord}{\mathcal{O}} 
  
 \newcommand{\lam}{\lambda} 
 \newcommand{\lamp}{\lambda'}

 \newcommand{\sgnmu}{\textrm{sgn}(\mu)} 
  
 \newbox\charbox  
 \newbox\slabox  
 \def\s#1{{      
     \setbox\charbox=\hbox{$#1$}  
     \setbox\slabox=\hbox{$/$}  
     \dimen\charbox=\ht\slabox  
     \advance\dimen\charbox by -\dp\slabox  
     \advance\dimen\charbox by -\ht\charbox  
     \advance\dimen\charbox by \dp\charbox  
     \divide\dimen\charbox by 2  
     \raise-\dimen\charbox\hbox to \wd\charbox{\hss/\hss}  
     \llap{$#1$}  
 }}

\begin{document} 

\title{Constraints on the R-Parity-violating minimal 
       supersymmetric standard model with neutrino masses from 
       multilepton studies at the LHC}

\author{M. Hanussek} 
\email[]{hanussek@th.physik.uni-bonn.de} 
\affiliation{Bethe Center for Theoretical Physics, Nu{\ss}allee 12, 53115 Bonn, Germany and SISSA/ISAS, Via Bonomea 265, I-34136 Trieste, Italy} 
 
\author{J.~S. Kim} 
\email[]{jongsoo.kim@adelaide.edu.au} 
\affiliation{ARC Centre of Excellence for Particle Physics at the Terascale, 
 School of Chemistry and Physics, University of Adelaide, SA 5005, Australia}

\preprint{ADP-12-44/T811, SISSA 30/2012/EP}

 \begin{abstract} 
   In a recent paper, we proposed a hierarchical ansatz for the 
   lepton-number-violating trilinear Yukawa couplings of the R-parity 
   violating minimal supersymmetric standard model. As a result, the 
   number of free parameters in the lepton-number-violating sector 
   was reduced from 36 to 6. Neutrino oscillation data fixes these six 
   parameters, which also uniquely determines the decay modes of 
   the lightest supersymmetric particle and thus governs the collider signature at the LHC.
   A typical signature of our model consists of multiple 
   leptons in the final state and significantly reduced missing 
   transverse momentum compared to models with R-parity conservation. In 
   this work, we present exclusion limits on our model based on 
   multilepton searches performed at the Large Hadron Collider with 7 TeV 
   center-of-mass energy in 2011 while accommodating a 125 GeV Higgs.
 \end{abstract} 

  \maketitle

 \section{Introduction} 
Recently, the Large Hadron Collider (LHC) experiments announced discovery of a 
Higgs-like particle with a mass around 125 GeV \cite{:2012gu,:2012gk}. To all appearances, the properties of this 
 Higgs-like particle seem to agree well with the standard model (SM) 
 Higgs boson. It is essential to probe whether the Higgs-like particle is 
 embedded in the SM or whether electroweak symmetry breaking is realized in 
 an extended framework of the SM. However, it is 
 also equally important to look for extensions of the SM in direct 
 collider searches for phenomena of physics beyond the 
 SM. Supersymmetry (SUSY) \cite{Haber:1984rc,Drees:2004jm} is a popular 
 extension of the SM and the lightest supersymmetric particle (LSP) is 
 most commonly stabilized by imposing R-parity conservation. Many SUSY
 searches at the LHC focus on collider signatures with a low SM 
 background by demanding large missing transverse momentum 
~\cite{Aad:2011qa,Chatrchyan:2011zy}. 
 So far, the searches for large 
 missing transverse momentum in association with multiple hard jets 
 provide the strictest bounds on supersymmetric models, where the 
 lightest neutralino is assumed to be the LSP. 
  
 In R-parity-violating models, the LSP is unstable and its decay 
 products may be detected within the detector. Thus less missing 
 transverse momentum is produced on average and it is interesting to 
 investigate how strict the resulting bounds are compared to the 
 R-parity conserving case.  The decay properties of the LSP are 
 determined by the R-parity-violating couplings.  A priori, there is a 
 large number of trilinear and bilinear R-parity-violating parameters. 
 However, simultaneous lepton and baryon number violation leads to 
 experimentally unobserved proton decay \cite{Dreiner:1997uz}.  As an 
 equally well motivated alternative to R-parity, one can impose the 
 discrete symmetry baryon triality ($\bt$) \cite{Dreiner:2005rd, 
   Dreiner:2012ae}, which forbids the baryon number violating terms but 
 allows for lepton-number violation. Then, massive neutrinos arise in 
 the mass spectrum \cite{Kaplan:1999ds,Allanach:2007qc} and the 
 neutrino oscillation data \cite{Schwetz:2011qt,An:2012eh} can be used 
 to constrain the lepton-number-violating sector. In a previous publication we proposed a 
 hierarchical structure of the R-parity-violating trilinear Yukawa 
 couplings similar to the SM Yukawa couplings \cite{Dreiner:2007uj}. As 
 a consequence, there remain only 6 independent lepton-number-violating 
 parameters, which can be completely fixed by the neutrino oscillation 
 data, uniquely determining the decay properties of the 
 LSP. Characteristic collider signatures contain multiple leptons, 
 third generation particles, several hard jets and, most importantly, reduced missing 
 transverse momentum. 
  
 In a recent study \cite{Hanussek:2012eh}, we examined the impact of 
 the three most important R-parity conserving searches from ATLAS 
 \cite{Aad:2011ib,ATLAS:2011ad,Aad:2011cwa} with missing 
 transverse momentum and multi-jets on our hierarchical R-parity 
 violating model, assuming a constrained soft-breaking sector (cMSSM) 
 \cite{Weinberg:1979sa,Sakai:1981pk,Weinberg:1981wj}. In the light of 
 the recent little hierarchy problem \cite{Hall:2011aa,Ghilencea:2012gz,Baer:2012mv}, 
the cMSSM seems to have become somewhat unnatural, since
in order to obtain a 125 GeV Higgs boson, large radiative corrections to the tree-level Higgs mass 
from scalar top quarks are needed, resulting in large fine-tuning in cMSSM models. However, a 125 GeV Higgs boson 
can still be obtained with percent level fine-tuning in the focus point region of a
(slightly modified) cMSSM \cite{Feng:2012jf} or in the focal curve region \cite{Akula:2011jx}; 
hence for simplicity we continue to present our results in the cMSSM framework.
 
 Both ATLAS and CMS have recently published multilepton searches where the cut on missing transverse 
 momentum is significantly reduced or even replaced by a cut on the scalar sum ($S_T$) of the transverse momentum of all 
 reconstructed objects~\cite{:2012kr,Chatrchyan:2012ye}. These 
 searches are expected to be more sensitive to our model than the R-parity conserving studies because the LSP decays reduce missing transverse
 momentum but not $S_T$.
 ATLAS interprets their null result in a simplified model of chargino pair production with
  a neutralino LSP decaying via a R-parity-violating coupling into 2 leptons and a neutrino and a scenario with a stau LSP candidate, 
  considering final state signatures with 4 or more leptons. In our framework, the LSP decay 
  modes are different due to the presence of neutralino-neutrino mixing terms, which tend
  to reduce the number of final state leptons. 
 Therefore, we find that the CMS study is more effective in setting exclusion limits on our model, because they also present signal regions with 3 or more leptons. Thus, we will in the following focus on the CMS study, presenting exclusion limits on our framework, while also taking into account a 125 GeV Higgs
in the focus point region. Our results can be interpreted in terms of bounds on the lighter chargino mass. 
    
 Our analysis is structured as follows: In 
 Sec.~\ref{sec:neutrinomass}, we shortly discuss how neutrino masses 
 arise in the hierarchical $\bt$ cMSSM and in Sec.~\ref{sec:numerics} 
 we outline our numerical procedure. In Sec.~\ref{sec:collider}, we 
 constrain the parameter space of the hierarchical $\bt$ cMSSM using 
 the multilepton ATLAS and CMS searches. We conclude in 
 Sec.~\ref{sec:conclusion}. 
  
 \section{Hierarchical Baryon Triality And Neutrino Masses} 
 \label{sec:neutrinomass} 
 The hierarchical $\bt$ MSSM allows for additional, L-violating terms 
 in the superpotential compared to the $\rp$ MSSM superpotential 
 ($\spot_{\rp}$)~\cite{Weinberg:1979sa,Sakai:1981pk,Weinberg:1981wj}, 
 \bea  
 \spot_{\bt} &=& \spot_{\rp} +  \frac{\lam_{ijk} }{2}
 L_i L_j \bar{E}_k + 
 \lamp_{ijk} L_i Q_j \bar{D}_k \nonumber \\
 & & - \kappa_i L_iH_u. 
 \eea  
 $L_i$, $Q_i$ denote the SU(2) doublet lepton and quark 
 superfields. $\bar E_i$, $\bar D_i$ are the SU(2) singlet lepton and 
 down-type quark superfields, respectively.  $i,j,k$ are 
 the family indices, while the $SU(2)_L$ and 
 $SU(3)_c$ indices are suppressed.  We work in a basis where the 
 bilinear R-parity-violating superpotential term and the corresponding 
 soft breaking term are rotated away by a field redefinition at the 
 unification scale \cite{Allanach:2003eb,Dreiner:2003hw}. Note that 
 these terms reemerge at lower scales via the renormalization group 
 equations. 
  
 The hierarchical ansatz implies that the trilinear L-violating 
 couplings have the following form~\cite{Dreiner:2007uj} 
 \begin{eqnarray} 
 \lam_{ijk}&\equiv&\ell_{i}\cdot \left(Y_E\right)_{jk}~-~\ell_{j}\cdot \left(Y_E\right)_{ik} 
 \label{ansatz_lam}\,, 
 \\ 
 \lam^\prime_{ijk}&\equiv&\ell^\prime_{i}\cdot \left(Y_D\right)_{jk} 
 \label{ansatz_lamp}\, 
 \end{eqnarray} 
 where $\ell_i,\,\ell^\prime_i$ are $c$-numbers. Eq.~(\ref{ansatz_lam}) 
 has the required form to maintain the anti-symmetry of the 
 $\lam_{ijk}$ in the first two indices. Assuming a specific form of the 
 Higgs-Yukawa couplings $Y_{E/D}$, the trilinear L-violating couplings 
 are fully determined by the 6 independent parameters $\ell_i,\,\ell^\prime_i$. 
  
 In the hierarchical $\bt$ constrained MSSM (cMSSM), the number of free 
 parameters in the soft-breaking sector is constrained to five, and 
 thus the model is described by 11 independent parameters at the 
 unification scale \cite{Allanach:2003eb}, 
 \beq 
  M_0,\,M_{1/2},\,A_0,\, \sgnmu,\,\tan{\beta},\,\ell_i,\,\ell^\prime_i.  
 \label{B3CMSSM} 
 \eeq 
 $M_0$, $M_{1/2}$ and $A_0$ denote the universal scalar mass, universal 
 gaugino mass and universal trilinear scalar coupling, 
 respectively. $\sgnmu$ is the sign of the superpotential Higgs mixing 
 parameter and $\tan\beta$ is the ratio between the two Higgs vacuum 
 expectation values.  
  
 Massive neutrinos emerge in the mass spectrum of the hierarchical 
 $\bt$ cMSSM because the neutrinos mix with the neutralinos via the 
 L-violating terms \cite{Allanach:2003eb}. However, it is well known 
 that at tree-level only one neutrino mass eigenstate obtains a mass 
 \cite{Allanach:2007qc}.  The global fit results to the neutrino 
 oscillation data \cite{Schwetz:2011qt,An:2012eh} show that we need at 
 least one further massive neutrino, which arises at one loop level.
 Full one loop contributions to the neutrino-neutralino mass matrix 
 are implemented in {\tt SOFTSUSY} \cite{Allanach:2011de}.  As 
 described in Refs. \cite{Dreiner:2010ye,Dreiner:2011ft}, the ratio 
 between the tree-level neutrino mass and the radiative contributions 
 is $\ord(100)$ in large regions of the cMSSM parameter space, 
 contradicting the experimental observation of a neutrino mass 
 hierarchy of $\ord(1)$.  However, neutrino masses with the correct 
 neutrino mass hierarchy can be obtained if the trilinear universal 
 scalar coupling is fixed to 
 \bea 
 A_0^{(\lambda^\prime)}\approx 2 M_{1/2}, 
 \label{A0_minimum_lp} 
 \eea 
 This approximation holds until $M_0 \gg2 M_{1/2}$. Then, 
$A_0^{(\lambda^\prime)} $ grows linearly with $M_0$.

 \section{Numerical procedure}\label{sec:numerics} 
 The low energy mass spectrum and couplings including neutrino masses 
 are calculated with {\tt SOFTSUSY3.3} \cite{Allanach:2011de}.  The 
 L-violating parameters $\ell_i$ and $\ell_i^\prime$ are determined by 
 a fitting procedure using the most recent experimental neutrino data 
 \cite{Hanussek:2012eh} with the root package {\tt MINUIT2}. We assume 
 $\tan\beta=25$ and the Higgs mixing parameter $\mu>0$, while $A_0$ is 
 determined by Eq.~(\ref{A0_minimum_lp}), leaving only $M_0$ and 
 $M_{1/2}$ as free parameters.  For the derivation of exclusion limits 
 on our model, we perform a scan in the $M_0$-$M_{1/2}$ plane. 
  
 The decay widths of the relevant sparticles are obtained with {\tt 
   IsaJet7.64}~\cite{Paige:2003mg} and {\tt IsaWig1.200}. Because 
 neutralino LSP decays via the sneutrino vevs and the $\kappa_i$ terms 
 are not implemented in {\tt IsaWig1.200}, we evaluate these with {\tt 
   SPheno3.1}~\cite{Porod:2011nf}.  We use the parton distribution 
 functions MRST2007 LO modified~\cite{Sherstnev:2007nd}. Our signal 
 events~\cite{foot1} are generated with {\tt 
   Herwig6.510}~\cite{Corcella:2000bw}. The cross sections are 
 normalized with the NLO calculations from {\tt Prospino2.1}~ 
 \cite{Beenakker:1996ed} assuming equal renormalization and 
 factorization scale.  We take into account detector effects by using 
 the fast detector simulation {\tt Delphes1.9} \cite{Ovyn:2009tx}.  Our 
 event samples are then analyzed with the program package {\tt ROOT} 
 \cite{Brun:1997pa} and we calculate the $95\%$ and $68\%$ confidence 
 levels (C.L.) of the exclusion limits by the Rolke test 
 \cite{Rolke:2004mj}.

 In the following section, we investigate the implications of 
 multilepton LHC searches for our model. 
  
 \section{Multi-Lepton Searches at the LHC} 
 \label{sec:collider} 
  
 In this section, we derive exclusion limits on the $M_0$-$M_{1/2}$ 
 plane in the hierarchical $B_3$ cMSSM from a recent multilepton 
 search at the LHC with 2011 data.  We discussed the collider 
 signatures of the hierarchical $B_3$ cMSSM in detail in 
 Ref.~\cite{Hanussek:2012eh}, for both the neutralino and the stau LSP 
 case.  There, we concluded that we expect 3 or more leptons from 
 neutralino LSP decays in $\ord(10 \%)$ of the events, and 2 leptons in up to 
 $30\%$ of events. Stau LSP decays result in 2 leptons in 40\% of events. 
 Apart from the LSP decays, the decay chains are identical to the R-parity 
 conserving case, since the L-violating couplings are quite small 
 ($\ord(10^{-5})$). 
 Within the decay chain, additional leptons from 
 e.g. chargino decays can arise, so that we expect a sizable fraction 
 of events with three or more leptons.  
  
 In the following, we discuss the CMS study
 \cite{Chatrchyan:2012ye} at $\sqrt{s}=7$ TeV with an integrated 
 luminosity of 4.98 fb$^{-1}$ and then present our numerical results 
 taking a 125 GeV Higgs into account.
  
 The CMS study is divided into many mutually 
 exclusive search channels with either 3 or 4 isolated leptons in the 
 final state (including hadronically decaying tau leptons) and for 
 three different $S_T$ regions ($S_T<300$ GeV, 300 GeV $<S_T<$ 600 GeV 
 and $S_T>$ 600 GeV). The CMS study uses the $S_T$
 of the transverse momentum of all reconstructed objects as a selection 
 cut. Since missing transverse momentum ($\met$) is not a powerful discriminating observable in 
 R-parity-violating models as opposed to R-parity conserving models, 
 the overall mass scale of the process is here certainly the more 
 appropriate observable. CMS also considers search channels with a mild 
 missing transverse momentum cut. However, these search channels yield much 
 weaker constraints on our R-parity-violating models compared to the 
 $S_T$ search channels. Events are further categorized according to the 
 number of Drell-Yan 
 pairs with either Z boson veto or acceptance.  Any new physics model 
 is allowed to have an excess in a limited amount of channels, which 
 are defined as signal region, and the remaining channels with no 
 excess are designated as control regions. 
  
\begin{center} 
 \begin{figure} 
 \centering 
 \includegraphics[scale=0.4]{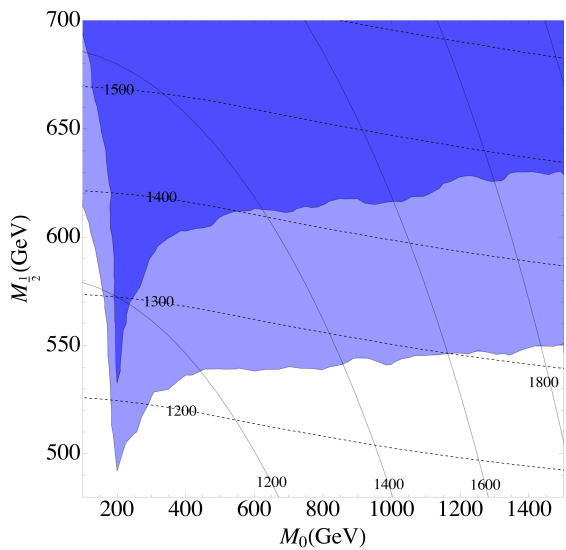} 
 \caption{Exclusion limit on our benchmark region, where 
   $\tan\beta=25$, $\sgnmu=1$ and $A_0^{(\lamp)} \approx 2 {M_{1/2}}$, 
  from the 3 lepton CMS study. The white region is excluded at 
   $95\%$ C.L., the light blue is excluded at $68\%$ C.L.. The grey lines 
   denote the squark masses, the dashed black lines denote the gluino 
   masses (each in GeV).} 
 \label{3lept_CMS} 
 \end{figure} 
 \end{center}

 In our case, the most promising channel is the 3 lepton channel with 
 $S_T>600$ GeV, without hadronically decaying taus and without 
 Drell-Yan pairs. This is because a strict $S_T$ cut suppresses the SM 
 background most effectively and because of the limited tau tagging 
 efficiency of Delphes \cite{Desch:2010gi}. Less than 
 $\ord(1\%)$ of our events contain Drell-Yan pairs from Z boson decays 
 (cf.~our discussion in Ref.~\cite{Hanussek:2012eh}); In 
 Fig.~\ref{3lept_CMS}, we present the exclusion limits in the 
 $M_0$-$M_{1/2}$ plane. We also show the contours of constant squark 
 and gluino masses, which range from 1 to 2 TeV.  Thus, the production 
 of colored states is heavily suppressed in large regions of parameter 
 space compared to the production of electroweak gauginos. The 
 lighter chargino is mostly wino-like  and typically chargino-neutralino 
 and chargino pair production 
are the dominant production channels for large $M_0$, $M_{1/2}$.
  
 The exclusion limits have a peak at small $M_0$ values. Here, the stau 
 is the LSP and slepton pair production yields a sizable contribution 
 to the total sparticle cross section besides colored sparticle 
 and chargino production. The stau LSP always decays in a two 
 body final state. As long as the channel $\tilde\tau^-\rightarrow b\, 
 \bar t$ is kinematically suppressed or closed, the dominant decay modes are 
 $\tilde\tau^-\rightarrow \ell^- \, \nu_{\ell}$. We can obtain additional leptons from the cascade decay 
 chain and thus we often have 3 isolated leptons in the final 
 state. With increasing $M_0$, the exclusion limits sharply drops. In 
 this region, the stau decay into a top and a bottom is kinematically 
 open and becomes dominant, strongly reducing the number of isolated 
 leptons in the final state. 
  
 For $M_0 \gtrsim 200$ GeV, the stau becomes heavier than the 
 neutralino. Initially, the three-body decay mode $\chi_1^0 
 \rightarrow \nu\, b \bar b$ dominates, resulting in poor efficiency of 
 the trilepton study.  Then, the exclusion limit moves again towards 
 higher values of $M_{1/2}$, since with increasing $M_0$ the two body 
 neutralino decay modes via bilinear L-violating couplings or 
 sneutrino vacuum expectation values quickly become dominant.  Thus, a 
 sizable fraction decays into W's and charged leptons. Additional 
 leptons arise from the lighter chargino decay into $W^\pm$ and 
 $\tilde\chi_1^0$. As mentioned before, squark and gluino production 
 decreases with increasing $M_0$ and $M_{1/2}$, such that electroweak gaugino pair 
 production is dominant for $M_0\gtrsim 400$ GeV. The electroweak gaugino pair 
 production cross section then increases slightly with $M_0$
 (for constant 
 $M_{1/2}\sim 550$ GeV), as destructive interference terms become
 more suppressed due to heavier sfermion masses. But for very large $M_0$ values, the chargino gets heavier and thus the cross section falls off again.
  
\begin{center} 
 \begin{figure} 
 \centering 
 \includegraphics[scale=0.4]{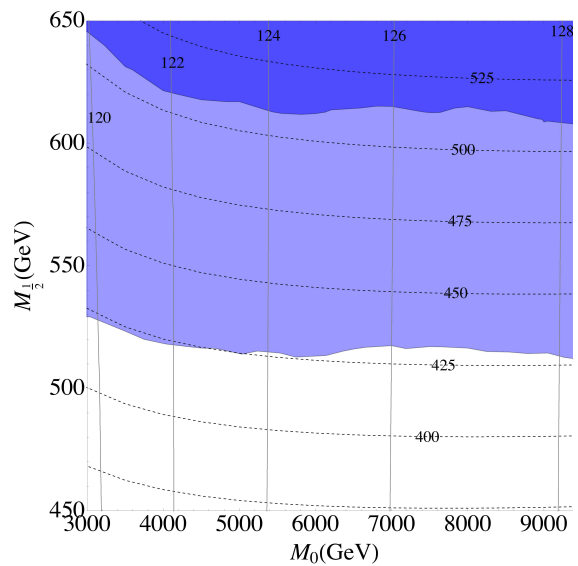} 
 \caption{Exclusion limit as in Fig.~\ref{3lept_CMS}, but for the focus point region, $M_0,\,A_0 \gg M_{1/2}$.
 The grey lines  denote the lightest Higgs mass, the dashed black lines denote the lighter chargino 
   mass (each in GeV).} 
 \label{3lept_CMS2} 
 \end{figure} 
 \end{center}

 We want to conclude the numerical discussion with commenting on the 
 Higgs mass in our scenario. As already mentioned, both experiments at CERN discovered 
 a 125 GeV resonance which is consistent with a SM-like Higgs. This 
 puts strong constraints on our model, since $A_0$ is fixed to be 
 positive and similar in magnitude to $M_{1/2}$ or $M_0$~\cite{foot2}.
 We obtain a Higgs mass of the order of $125\pm3$
 GeV only if $M_0  \gtrsim 5$ TeV for $M_{1/2}\sim 550$ GeV, as displayed in Fig.~\ref{3lept_CMS2}.
 In this region of parameter space, 
 the only sub-TeV sparticles are neutralinos and charginos,
 whereas the gluinos have a mass of $\sim$1.4 TeV. The scalars with masses around 3-6 
 TeV are beyond the reach of the current run of the LHC.

The region $M_{1/2}\ll M_0,\,A_0$ is similar to 
 the focus point region with large $A$-terms studied in Ref. \cite{Feng:2012jf}.
 Even though heavy scalars are disfavored by fine-tuning studies, we can still have small fine-tuning of one percent in the focus point region. However, the soft breaking sector of the cMSSM must be slightly modified \cite{Feng:2012jf,focus_neutrino}, allowing for $\ord{(10\%)}$ deviations in the stop soft-breaking terms.
 Note that demanding radiative electroweak symmetry breaking sets a relatively weak lower bound 
 on $M_{1/2}$ of roughly 50 (200) GeV for $M_0=5(10)$ TeV (recall that 
 for $M_0 \gg M_{1/2}$,  viable neutrino masses require $A_0\sim M_0$).

     
 \section{Conclusion} 
 \label{sec:conclusion} 
 We considered a hierarchical ansatz for the L-violating trilinear 
 Yukawa couplings in the $\bt$ cMSSM which enables us to unambiguously 
 determine the numerical values of all L-violating couplings from experimental 
 neutrino data. An important feature of this ansatz is that it 
 uniquely fixes the decay channels of the LSP and thus the collider 
 signature at the LHC. 
In the light of the 125 GeV Higgs and neutrino oscillation data, the hierarchical $B_3$ cMSSM predicts heavy squarks and sleptons beyond the reach of the LHC, as well as normal mass ordering in the neutrino sector. This seems to be consistent with null results from direct collider searches at the LHC. Naively, one would expect large fine-tuning due to the heavy stop masses. However, our parameter space is similar to the focus point region with large $A$-terms resulting in less fine-tuning if one allows for $\ord(10\%)$ deviations in the third generation squark sector. In our model, electroweak gaugino pair production is the dominant sparticle production process. We confronted our model with a multilepton search 
 from CMS. We showed that the three lepton final state with 
 high $S_T$ and no $\met$ requirement provides the best exclusion 
 limits. We demonstrated that the CMS multilepton search with data 
 collected in 2011 already excludes large regions of the parameter 
 space up to values of $M_{1/2}$ close to 550 GeV. This bound is mostly
 independent of $M_0$, corresponding to bounds on wino-like charginos and neutralinos with masses around 430 GeV and squark and gluino 
   masses well above 1 TeV. Previously obtained exclusion limits from 
   $\met$ and multi-jet studies with 2011 data are much weaker, as 
   described in a previous publication, confirming that the use of 
   $S_T$ is a very powerful observable for R-parity-violating models. 

 \medskip 
  
 {\bf Acknowledgements.}  This work is supported in part by the ARC 
 Centre of Excellence for Particle Physics at the Terascale, the 
 Deutsche Telekom Stiftung and by the Bonn-Cologne Graduate School of 
 Physics. We thank M. Drees and C.H. Kom for useful discussions.


\begin{thebibliography}{99}
\bibitem{foot1} 
  Pair production processes of all SUSY particles have been simulated.
\bibitem{foot2}
  The former case is applicable for $M_0\lesssim 2M_{1/2}$, whereas the latter case holds for $M_0\gg2M_{1/2}$.
\end{thebibliography}
 \end{document}